\documentstyle[12pt,aaspp4,eqsecnum,flushrt]{article}

\newcommand{\etal}{et al.}
\newcommand{\mic}{\mbox{\,$\mu$m}}      
\newcommand{\sM}{\mbox{\,s$^{-1}$}}      
\newcommand{\cmMM}{\mbox{\,cm$^{-2}$}}      
\newcommand{\cmMMM}{\mbox{\,cm$^{-3}$}}      
\newcommand{\ergcmM}{\mbox{\,erg\,cm$^{-1}$}}      
\newcommand{\gcmMMM}{\mbox{\,g\,cm$^{-3}$}}      
\newcommand{\be}{\begin{equation}}
\newcommand{\ee}{\end{equation}}
\newcommand{\ba}{\begin{eqnarray}}
\newcommand{\ea}{\end{eqnarray}}
\newcommand{\Td}{\mbox{$T_{\rm d}$}}
\newcommand{\Tg}{\mbox{$T_{\rm g}$}}
\newcommand{\Teff}{\mbox{$T_{\rm eff}$}}
\newcommand{\HHO}{\mbox{H$_2$O}}
\newcommand{\COO}{\mbox{CO$_2$}}
\newcommand{\HH}{\mbox{H$_2$}}
\newcommand{\Rvap}{\mbox{$R_{\rm vap}$}}      
\newcommand{\Rvaprt}{\mbox{\Rvap$(r,t)$}}      
\newcommand{\Pvap}{\mbox{$P_{\rm vap}$}}      
\newcommand{\Pvaprt}{\mbox{\Pvap$(r,t)$}}      
\newcommand{\Nvap}{\mbox{$N_{\rm vap}$}}      
\newcommand{\tprime}{\mbox{$t^{\prime}$}}      
\newcommand{\Aeff}{\mbox{$A_{\rm eff}$}}      
      
\newcommand{\vecA}{\mbox{{\boldmath$A$}}}      
\newcommand{\vecJ}{\mbox{{\boldmath$J$}}}

\newcommand{\vecz}{\mbox{{\boldmath$z$}}}

\newcommand{\barJk}{\mbox{$\overline{\delta J_k}$}}      
\newcommand{\barJkJl}{\mbox{$\overline{\delta J_k\delta J_l}$}}

\newcommand{\vvap}{\mbox{$v_{\rm vap}$}}
\newcommand{\vth}{\mbox{$v_{\rm th}$}}

\newcommand{\bfzh}{\mbox{$\hat{\vecz}$}}

\newcommand{\EGeV}{\mbox{${E \over {\rm GeV}}$}}      
\newcommand{\Zfac}{\mbox{$\left({Z \over 26}\right)^2$}}      
\newcommand{\mvap}{\mbox{$m_{\rm vap}$}}      
\newcommand{\Tvap}{\mbox{$T_{\rm vap}$}}      
\newcommand{\tmax}{\mbox{$t_{\rm max}$}}      
      
\newcommand{\thets}{\mbox{$\theta_{\rm s}$}}      
\newcommand{\tdamp}{\mbox{$t_{\rm damp}$}}      
\newcommand{\Bkkg}{\mbox{$B_{kk}^{\rm (g)}$}}      
\newcommand{\Bkkcr}{\mbox{$B_{kk}^{\rm (rck)}$}}      
      
\newcommand{\Iunit}{\mbox{cm$^{-2}$\,s$^{-1}$\,GeV$^{-1}$}}      

\newcommand{\Erot}{\mbox{$E_{\rm rot}$}}

\begin{document}
\title{Cosmic rays and grain alignment}
\author{A. Lazarian\altaffilmark{1} and W. G. Roberge\altaffilmark{2} }
\altaffiltext{1}{Dept.\ of Astrophysical Sciences, Princeton University,
Princeton, NJ 08544, USA}
\altaffiltext{2}{Dept.\ of Physics, Applied Physics \& Astronomy,
Rensselaer Polytechnic Institute, Troy, NY 12180, USA}

\begin{abstract}
The recent detection of interstellar polarization in the solid CO
feature near 4.67\mic\ shows that CO-mantled grains can be aligned
in cold molecular clouds.
These observations conflict with a theory of grain alignment
which attributes the polarization in molecular clouds to
the effects of cosmic rays: according to this theory, oblate spheroidal
grains with \HHO- and \COO-dominated ice mantles are spun up to
suprathermal energies by molecular evaporation from cosmic ray
impact sites but spin up does not occur for CO-mantled grains.
Motivated by this conflict, we reexamine the effects of
cosmic rays on the alignment of icy grains.
We show that the systematic torques produced by cosmic rays
are insufficient to cause suprathermal spin.
In principle, the random torques due to cosmic rays
can enhance the efficiency of Davis-Greenstein alignment by raising
the grain rotational temperature.
However, a significant enhancement would require cosmic
ray fluxes 6--7 orders of magnitude larger than the flux
in a typical cold cloud.
\end{abstract}

\setcounter{footnote}{0}

%
%
\section{Introduction}

The recent detection of polarization in the 4.67\mic\
feature of solid CO (Chrysostomou \etal\ 1996) sets severe
constraints on theories of interstellar grain alignment.
Solid CO mantles are expected to survive only in cold, dense
clouds, where the temperatures of the gas (\Tg) and grain solid
material (\Td) must be nearly equal due to collisional coupling.
For example, the Davis-Greenstein alignment mechanism is a
dissipative process which is driven by the difference between $\Tg$ and $\Td$
(Jones \& Spitzer 1967).
Recent calculations (DeGraff,
Roberge \& Flaherty 1997) show that the largest polarizations observed
toward molecular clouds are inconsistent with 
Davis-Greenstein alignment unless
$\Tg/\Td \gtrsim 10$ and that this conclusion holds even if the
grains are superparamagnetic.\footnote{However, helical grains may
be aligned even if $\Tg\approx \Td$ (see Lazarian 1995).
The process is discussed at length in Lazarian, Goodman \& Myers (1996).}
Since a temperature disparity of this magnitude is inconsistent
with observations, it seems certain that the grains in cold
clouds--- including CO-mantled grains--- are aligned by
some ``nonthermal'' process.

The first nonthermal model of grain alignment was developed
by Purcell (1975, 1979, hereafter P79), who pointed out that
the temperatures of the gas and dust do not limit the efficiency of
alignment if the grain rotational energies are sufficiently large.
P79 pointed out that spin up to suprathermal energies will occur
whenever the torque on a grain, referred to axes fixed in the grain material,
has a nonvanishing time average over times $\gtrsim$ the timescale
for frictional coupling to the gas.
P79 also identified three mechanisms that can produce
such a ``pinwheel torque'':
(i) the formation of H$_2$ at catalytic sites on the grain surface, which
    requires atomic hydrogen;
(ii) gas-grain collisions on a surface with spatial variations
     in the ``accommodation coefficient,'' which requires $\Tg\neq\Td$;
     and
(iii) photoelectric emission from a surface with variations in
     the photoelectric yield, which requires UV photons.
All of these processes are suppressed in dark clouds, where the atomic
hydrogen concentration and UV flux are negligible and $\Tg\approx\Td$.
If the starlight flux in the blue part of the
spectrum\footnote{For ``standard'' grains of size
$\sim 10^{-5}$\,cm, only short-wavelength radiation causes an appreciable
torque (Draine \& Weingartner 1996a). However, if the aligned grains in
molecular clouds are substantially larger, then longer wavelengths will 
affect the alignment.}
is negligible, then we may also disregard the radiative torques arising
from differential scattering (Dolginov \& Mytrophanov 1975, 
Lazarian 1995a, Draine \& Weingartner 1996a,b).

In view of these circumstances, it is natural to inquire
whether some other process might cause suprathermal rotation.
A natural possibility is related to the
evaporation of molecules adsorbed on grain mantles:
if a grain is not heated uniformly, then hotter places on
the mantle will desorb molecules at higher
rates compared to colder places.  These ``hotspots'' can act
in a manner that is similar to the catalytic sites of H$_2$ formation
in Purcell's model for suprathermal spin up.
The required temperature nonuniformity might be caused
by nonuniformities in the absorption of light.
Consider, for example, a small soot inclusion on the grain surface.
Such an inclusion would absorb light and have a temperature greater than
the mean temperature of the grain. In other circumstances, a
darker inclusion would radiate more efficiently
and therefore become colder than the rest of the grain.
Since variations in the grain
absorption (emissivity) are likely to persist for much more than
a gas damping time, the resulting torques should be
long lived. If the grain dynamics are dominated by these
torques, rather than the torques due to gas damping,\footnote{For
this purpose, the randomization of angular momentum during a spin-up interval
should be small (Lazarian 1995b).} the alignment
could be substantial. We plan to address this interesting possibility
in a subsequent paper.

In this paper, we examine the feasibility of a related
mechanism suggested by Sorrell (1995a,b).
Sorrell pointed out that a cosmic ray impact heats a grain locally,
and that molecular evaporation from the resulting hot spot
will cause rotational acceleration via the rocket effect.
According to Sorrell's analysis, the rocket effect on an individual
grain produces a nonzero time-average torque with a correlation time
that is limited by changes in the mantle surface.
In this view, the rocket effect spins up the grain to a
suprathermal kinetic energy and the alignment occurs 
via Purcell's mechanism (P79).
However, the scenario considered by Sorrell (1995a) 
conflicts with the observations of Chrysostomou et al.\ (1996):
Sorrell's model predicts that grains with \HHO- and \COO-dominated
mantles will spin up to energies $\sim 100 k\Tg$ but that spin up of
CO-mantled grains does not occur.
In an effort to understand this conflict,
we reexamine the effects of cosmic rays on the alignment of
ice-mantled grains.
In \S2, we model the evaporation of molecules from a
cosmic ray hotspot.
In \S3, we consider the conditions under which evaporating
molecules may produce a pinwheel torque and hence suprathermal rotation.
In \S4, we describe the possible effects of the {\it random}\/
torques due to evaporation and the resulting enhancement of
Davis-Greenstein alignment.
We summarise our results in \S5.

%
%
\section{Evaporation from a cosmic ray hot spot}

The evaporation of molecules caused by cosmic ray heating has been
discussed elsewhere (Watson \& Salpeter 1972;
de Jong \& Kamijo 1973;
Aannestad \& Kenyon 1979; L\'eger, Jura \& Omont 1985, henceforth LJO85).
Here we merely apply the results of LJO85 to model the
evaporation of \HH\ and CO molecules from \HHO-ice mantles.
A cosmic ray nucleus with charge $Ze$ and energy $E$ loses
energy in solid \HHO\ at a rate
\be
Q(Z,E) = \left\{
\begin{array}{ll}
9.40 \times 10^{-4}\,\Zfac\,\left(\EGeV\right)^{-0.75} \ergcmM,
      & 0.02 < \EGeV < 0.2 \\
      &                    \\
1.48 \times 10^{-3}\,\Zfac\,
      \left[1+0.1\,\left(\EGeV\right)^{-1.5}\right] \ergcmM,
      & 0.2 < \EGeV < 10 \\
\end{array}
\right. 
\label{2.1}
\ee
(LJO85).\footnote{The energy loss rates in eq.\ (\ref{2.1}) include the
corrections derived by LJO85 to account for oblique cosmic ray impacts
and partial escape from the grain of electrons ejected by the cosmic ray.
We have taken the density of \HHO\ ice to be $1$\gcmMMM.}
The energy is deposited initially (on a timescale
$\sim 10^{-11}$\,s) in a cylinder of radius
$r_0 \sim 50$\,\AA\ around
the cosmic ray track and spreads laterally thereafter by thermal diffusion.
Over the temperature range of interest here, the
thermal diffusivity of \HHO\ ice depends only weakly on
temperature (Zeller \& Pohl 1971).
Consequently, the heating of the mantle is well described
by the heat diffusion equation with constant thermal diffusivity.
The solution for the energy density, $U$, is
\be
U(r,t) = {Q \over 4\pi\alpha \left(t+t_0\right)}\,
         \exp\left[-{r^2 \over 4\alpha\left(t+t_0\right)}\right],
\label{2.2}
\ee
where $r$ is distance perpendicular to the cosmic ray track,
$t$ is time, $\alpha$ is the thermal diffusivity, and
\be
t_0 \equiv {r_0^2 \over -4\alpha \ln\left(1-f\right)}.
\label{2.3}
\ee
Here we have adopted somewhat arbitrary initial conditions,
such that $U(r,0)$ is a Gaussian function of $r$
with a fraction $f$ of the total energy contained
in $r<r_0$ at $t=0$.
In the following discussion, we will assume that $r_0=50$\,\AA\ and
set $f=0.5$.
The temperature, $T$, can be found from the relation
\be
U(r,t) = \int_{T_0}^{T(r,t)}\,\rho\,C_V\left(T^{\prime}\right)
\,dT^{\prime},
\label{2.4}
\ee
where $T_0$ is the temperature before the impact and
$\rho\,C_V$ is the volume specific heat.
In the calculations discussed below, we set $T_0=15$\,K
and adopt the values of $\alpha$ and $\rho\,C_V$ recommended by
LJO85 for \HHO\ ice.

We assume that molecules evaporate from a hotspot at the
classical rate,
\be
\Rvaprt = \nu_0\,\exp\left[-\Delta H_{\rm s}/kT(r,t)\right],
\label{2.5}
\ee
where $\nu_0$ and $\Delta H_{\rm s}$ are respectively the
lattice vibration frequency and binding energy of a molecule
adsorbed on the mantle surface.
The probability that a molecule at radius $r$ has evaporated before time
$t$ is therefore
\be
\Pvaprt = 1 -
\exp\left[-\int_0^t\,\Rvap\left(r,\tprime\right)\,d\tprime\right].
\label{2.6}
\ee
The mean value of the total number of molecules evaporated by a
single cosmic ray with charge $Z$ and energy $E$ is
\be
\Nvap(Z,E)  = \Aeff(Z,E)\,\thets,
\label{2.7}
\ee
where \thets\ (\cmMM) is the surface coverage before
the impact and
\be
\Aeff(Z,E) \equiv \int_0^{\infty}\,dr\,2\pi r\, \Pvap(r,\tmax)
\label{2.8}
\ee
is the effective area ``cleaned off'' by the cosmic ray.
We assume that no evaporation occurs for times $t>\tmax$,
where $\tmax(r)$ is the time when the mantle at radius $r$
has cooled to temperature $T_0$;
because virtually no evaporation occurs for $T \approx T_0$,
the arbitrariness of this definition has virtually
no effect on our numerical results.
The value of \Aeff\ depends on $Z$ and $E$ only through
the energy loss rate, $Q$.
The dependence of \Aeff\ on $Q$ is illustrated
in Figure~1 for the evaporation of
\HH\ ($\nu_0=7.5\times 10^{12}$, $\Delta H_{\rm s}/k=555$\,K; Sandford
\& Allamandola 1993) 
and CO ($\nu_0=7.0\times 10^{14}$, $\Delta H_{\rm s}/k=1030$\,K; LJO85)
molecules from \HHO\ ice.

The temperature of the vapor at a hot spot 
\be
\Tvap(Z,E) \equiv
{
\int_0^{\infty}\,dr\,2\pi r \, \int_0^{\infty}\,dt
\left[1-\Pvaprt\right]\,\Rvaprt\,T(r,t)
\over
\int_0^{\infty}\,dr\,2\pi r \, \int_0^{\infty}\,dt
\left[1-\Pvaprt\right]\,\Rvaprt\
}
\label{2.15}
\ee
is the mean (for all evaporating molecules) temperature of the surface at
the instant of evaporation.
The numerical evaluation of expression (\ref{2.15}) for CO
and \HH\ evaporations shows that, for both \HH\ and CO,
\Tvap\ increases from $\approx 50$\,K to $\gtrsim 200$\,K
as $Q$ increases from $10^{-4}$\,erg\,cm$^{-1}$ to
$10^{-2}$\,erg\,cm$^{-1}$.

%
%
\section{Constraints on the pinwheel torque due to cosmic rays}

The evaporation of a molecule from the mantle surface
produces an impulsive change in the grain angular
momentum, \vecJ, due to the rocket effect.
The rotational dynamics of a grain subject to many such
impulses are determined by the mean torque,
\be
\vecA \equiv \left<{\Delta\vecJ \over \Delta t}\right>,
\label{2.9}
\ee
and diffusion tensor,
\be
B \equiv \left<{\Delta\vecJ \Delta\vecJ \over \Delta t}\right>,
\label{2.10}
\ee
where $\Delta\vecJ$ is the cumulative change in \vecJ\ caused by
impacts during a time interval $\Delta t$ and the angle
brackets denote time averages.
The mean torque due to the rocket effect has components
\be
A_k^{\rm (rck)} = \frac{1}{2}\,S_{\rm d}\,\sum_Z\,\int_{E_{\rm min}}^{\infty}
      dE\,\phi_Z(E)\,\Nvap(Z,E)\,\overline{\delta J_k},
\label{2.11}
\ee
where $S_{\rm d}$ is the grain surface area,  $\phi_Z(E)\,dE$
is the omnidirectional flux of cosmic rays with charge
$Ze$ and energies between
$E$ and $E+dE$, and $\overline{\delta J_k}$ is the mean
angular impulse due to a single evaporating molecule.
In deriving expression (\ref{2.11}), we have assumed that the
flux of cosmic rays is isotropic in the grain frame and that each
cosmic ray impact creates 2 hotspots.
The analogous expression for the diffusion tensor is
\be
B_{kl}^{\rm (rck)} =
\frac{1}{2}\,S_{\rm d}\,\sum_Z\,\int_{E_{\rm min}}^{\infty}
      dE\,\phi_Z(E)\,\Nvap(Z,E)\,\overline{\delta J_k\delta J_l}.
\label{2.12}
\ee

The quantities \barJk\ and \barJkJl\ are determined by
averaging the angular impulse due to a single evaporation
over the position of the evaporation site and momentum
of the evaporating molecule.
Consequently, \barJk\ depends in general on the grain shape.
In Appendix~A, we show that $\barJk=0$ for an oblate 
spheroid, the shape adopted by Sorrell (1995a,b).
It follows that $A_{k}^{\rm (rck)}=0$, that is, there is no pinwheel
torque due to cosmic ray impacts on an oblate spheroid.
This statement is true for any surface of revolution.
 
Of course, real interstellar grains are not surfaces of
revolution and it is possible to show that pinwhell torques
do not vanish for less symmetric shapes (e.g. they do not vanish
for square prisms).
Nevertheless, we can place an upper limit on the magnitude
of the pinwheel torque for any shape by assuming, unrealistically,
that all of the angular impulses produced by evaporating molecules
lie along the same direction in the grain frame.
Since the cosmic ray hits on a real grain are uniformly distributed over
its surface, this assumption is obviously extremely optimistic.

Let $\overline{\delta J} = b\left(\mvap k\Tvap\right)^{1/2}$ be the mean
magnitude of an individual impulse, where $b$ is some 
characteristic linear dimension of our hypothetical grain.
In our optimistic scenario, the mean torque due to
the rocket effect would have magnitude
\be
A^{\rm (rck)} =  
\frac{1}{2}\,S_{\rm d}\,\thets\,\sum_Z\,\int_{E_{\rm min}}^{\infty}
 dE\,\phi_Z(E)\,\Aeff(Z,E)\,\overline{\delta J}(E).
\label{3.1}
\ee
Now suppose that all of the angular impulses lie along the
$k$th principal axis of inertia.
If we assume that the rotational friction is provided by
gas damping, then the grain would spin up to kinetic
energy \Erot, such that
\be
{E_{\rm rot} \over \frac{1}{2}kT_{\rm g}}
=
{\left[A^{\rm (rck)}t_{{\rm gas},k}\right]^2 \over I_kkT_{\rm g}},
\label{3.2}
\ee
where $I_k$ and $t_{{\rm gas},k}$ are respectively the
rotational inertia and gas damping time for rotation about
axis $k$.

We have evaluated expression (\ref{3.2}) using the 
functional form of $\phi_Z$ given by LJO85,
cosmic ray abundances given in Simpson (1983)
for the local ISM, and
typical values for the grain properties
($b=10^{-5}$\,cm, $\Td=15$\,K,
$\rho_{\rm d}=3$\,\gcmMMM) and physical conditions
($n_{\rm g}=10^4$\,\cmMMM, $\Tg=15$\,K) in the gas.
We find the same value, $E_{\rm rot} \sim 10^{-5}\, kT_{\rm g}$,
whether we assume that the evaporating molecule is \HH\
or CO.
Note that this is the {\it average}\/ excess energy for an ensemble
of grains. 
For the cosmic ray fluxes adopted here, most of the grains in the
ensemble actually experience no cosmic ray hits during the
interval $t_{{\rm  gas},k}$, so the average energy is somewhat
misleading.
However, we estimate that  the largest energies in the ensemble
are only $\sim 10^{-3}$k\Tg, so this does not alter our conclusion.

Evidently, cosmic rays cannot spin up the grains
under the most optimistic scenario that one can conceive.
It is easy to understand this result if one notes that
only heavy cosmic rays have energy loss rates sufficiently
large to produce a significant number of evaporations.
For example, the cosmic ray spectrum calculated by LJO85
peaks at $E \approx 0.3$\,GeV for both protons and iron
nuclei.
At $E=0.3$\,GeV, the energy loss rates for protons and
iron nuclei are
$Q_{\rm p} = 4 \times 10^{-6}$\ergcmM\
and
$Q_{\rm Fe} = 2 \times 10^{-3}$\ergcmM, respectively.
According to Figure~1, evaporations caused by protons impacts
would be completely negligible at this energy.

\section{Rotational excitation by random cosmic ray torques}

The stochastic torque
produced by the rocket effect can enhance Davis-Greenstein
alignment by increasing the grain rotational temperature
(Salpeter \& Wickramsinghe 1969; Purcell \& Spitzer 1971).
The random angular impulses produced by evaporating molecules
cause the angular momentum to change in random walk fashion so that,
in the absence of other processes, the $k$th component of \vecJ\
would increase without limit as
$J_k \propto \left[\Bkkcr t\right]^{1/2}$.
In reality, the rotational friction produced by gas damping  and
other dissipative processes limits the growth to
$J_k \approx \left[\Bkkcr\tdamp\right]^{1/2}$, where \tdamp\ is the
relevant damping time.
For example,
suppose that the only interactions of the grains with their environment
are provided by gas damping plus the rocket effect.
Then one can show (see RDGF93, eq.\ [3.18]) that the
distribution of $J_k$ is Maxwellian with an effective temperature
\be
\Teff =
{
\left[\, B_{kk}^{\rm (g)} + B_{kk}^{\rm (rck)} \,\right] t_{{\rm gas},k}
\over
2I_k k
},
\label{3.3}
\ee
where \Bkkg\ is the diffusion tensor for gas damping.
If \Bkkcr\ is sufficiently large, then the rotational
excitation provided by the cosmic rays can permit
Davis-Greenstein alignment even in clouds with $\Td=\Tg$.

It follows from equation (\ref{3.3}) that the stochastic
torque produced by the rocket
effect increases the grain rotational temperature
by a factor $1+\Theta$, where
\be
\Theta \equiv {B_{kk}^{\rm (rck)} \over B_{kk}^{\rm (g)}}.
\label{3.4}
\ee
After combining expressions (\ref{2.7}), (\ref{2.12}),
and (\ref{2.14}), we find that the nonzero components
of the diffusion tensor for the rocket effect are
\be
\Bkkcr = {4\pi \over 3}\,\Gamma_k\,
       m_{\rm vap}k\,b^4\,\thets\,\sum_Z W_Z,
\label{3.5}
\ee
where 
\be
W_Z \equiv \int_{E_{\rm min}}^{\infty}
      dE\,\,\phi_Z(E)\,\Aeff(Z,E)\,\Tvap(Z,E)
\label{3.6}
\ee
and the $\Gamma_k$ are weak functions of the grain
eccentricity with $3/8 \le \Gamma_k \le 1$ (see Appendix~A).
The diffusion tensor for gas damping is also diagonal
with components
\be
\Bkkg = {2\sqrt{\pi} \over 3}\,\Gamma_k\,
       n_{\rm g}m_{\rm g}^2b^4v_{\rm th}^3\,
       \left(1+{T_{\rm d} \over T_{\rm g}}\right)
\label{3.7}
\ee
(RDGF93),
where $n_{\rm g}$ is the number density of the gas,
$m_{\rm g}$ is the mass of a gas particle,
and $\vth \equiv \sqrt{2kT_{\rm g}/m_{\rm g}}$ is the gas thermal speed.
It follows that
\be
\Theta  =
{\sqrt{\pi} } \left(1+T_{\rm d}/T_{\rm g}\right)\,
\left({m_{\rm vap} \over m_{\rm g}}\right)\,
{\sum_Z W_Z \over n_{\rm g} v_{\rm th} \theta_{\rm s}^{-1} T_{\rm g} }.
\label{3.8}
\ee
Notice that $\Theta$ is independent of the surface eccentricity.

In order to estimate $\Theta$, we used the results of \S2
to evaluate expression (\ref{3.6}) by numerical integration.
Due to the steep dependence of \Aeff\ on $Q$ (Fig.\ 1), the sum is
dominated by iron, the most abundant nucleus with $Z\gg 1$.
For example, consider the relative contributions of iron and
cosmic ray protons.
LJO85 found that iron and protons have approximately the same energy
spectrum; according to their models, $\phi_Z = 4\pi A_Z I$, where
$A_Z$ is the abundance of cosmic rays with charge $Ze$ relative to
the abundance of protons and
\be
I(E) = \left\{
\begin{array}{lcl}
21\,\left(\EGeV\right)          & \ \ \ \ &  0.02 < \EGeV < 0.07    \\
1.5                             & \ \ \ \ &  0.07 < \EGeV < 0.2     \\
0.3\,\left(\EGeV\right)^{-1}    & \ \ \ \ &  0.2  < \EGeV < 1       \\
0.3\,\left(\EGeV\right)^{-2}    & \ \ \ \ &  1    < \EGeV           \\
\end{array}
\right.
\label{3.9}
\ee
(\Iunit) is the mean intensity of cosmic ray protons.
Taking $A_{26}=3\times 10^{-5}$, the cosmic ray abundance of iron
in the local interstellar medium (Simpson 1983 and references therein),
we find that
$W_{26}=5\times 10^{-13}$\,K\,\sM\ and
$W_1=4\times 10^{-14}$\,K\,\sM\ for
the evaporation of \HH\ from \HHO\ ice.
The analogous calculations for CO evaporation yield
$W_{26}=3\times 10^{-13}$\,K\,\sM\  and
$W_{1}=6\times 10^{-17}$\,K\,\sM.
We will assume henceforth that all of the evaporation is
caused by iron cosmic rays and set
$\sum_Z W_Z =4\times 10^{-14}$\,K\,\sM\ for
\HH\ evaporation and
$\sum_Z W_Z =5\times 10^{-13}$\,K\,\sM\ for \HH\ evaporation and
$\sum_Z W_Z =3\times 10^{-13}$\,K\,\sM\ for CO
evaporation.\footnote{To estimate the constributions
of other nuclei, we evaluated $\sum_Z W_Z$ by setting
$\phi_Z = 4\pi A_Z I$, assuming that $I$ is given by expression (\ref{3.9}),
and adopting cosmic ray abundances appropriate for the
local ISM (Simpson 1983).
In this approximation, adding the effects of nuclei with
$Z\ne 26$ increases $\sum_Z W_Z$ by less than a factor of
two for \HH\ and CO evaporation.
An error of this magnitude in $W_Z$ has no effect on our
conclusions.}
To estimate $\Theta$ for typical cold cloud conditions, we will
assume that the gas is composed of \HH, with $n_{\rm g}=10^4$\cmMMM,
$\Tg=\Td=15$\,K, and $\thets=10^{15}$\cmMM.
With these assumptions, we find that
$\Theta \sim 10^{-7}$ if the rocket effect is caused by the 
evaporation of \HH\ and
$\Theta \sim 10^{-6}$ if the evaporating molecules are
CO.
We conclude that the rotational excitation provided by the rocket effect
is insignificant unless the flux of iron nuclei were increased
by 6--7 orders of magnitude.
However, an increase in $\phi$ of this magnitude can
be ruled out in general: with the flux adopted in our estimates,
iron nuclei would already contribute about $10$\%
of the total cosmic ray ionization rate, $\zeta$, in a typical cloud
with $\zeta \sim 10^{-17}$\,\sM\ (LJO85).

In principle, the desorption produced by low-energy
electrons (Johnson 1990) could also contribute to the
rotational excitation.
However, the required electron flux is ruled out unambiguously
by data on the interstellar ionization.

Purcell \& Spitzer (1971) computed the rotational excitation
which is associated with energy loss by cosmic rays in the grain
solid material.
Comparing our results with those of Purcell \& Spitzer (1971), we
conclude that energy loss is more important for enhancing the
rotational temperature of grains than the desorption
of molecules from hotspots. Nevertheless, even the
process studied by Purcell \& Spitzer has only a marginal effect
on grain alignment.

%
%
\section{Summary}

Our study has shown that the evaporation of molecules from cosmic
ray hotspots produces no pinwheel torque on an oblate spheroidal grain,
regardless of its mantle composition.
Although it is possible, in principle, that nonzero pinwheel torques
exist for some grain shapes, we have shown that the resulting
increase in the grain rotational energy is negligible for realistic
fluxes of heavy cosmic rays.
Our findings
resolve the apparent conflict between the recent observations of
Chrysostomou et al.\ (1996) and the grain alignment model of
Sorrell (1995a), inasmuch as the alignment process postulated in Sorrell's
model does not occur.
We have also shown that the random torques caused by cosmic ray
hotspots produce rotational excitation that can
enhance the efficiency of Davis-Greenstein alignment,
even in clouds where $\Td=\Tg$.
However, the enhancement is completely insignificant for reasonable
estimates of the cosmic ray flux and less than the enhancement
associated with cosmic ray energy losses in the grain
material (Purcell \& Spitzer 1971).

\acknowledgements
We are grateful to Bruce Draine for valuable comments.
A.L. acknowledges the support of NASA grant NAG5~2858 and
W.G.R. acknowledges the support of NASA grant NAGW-3001.

%
%

\appendix
\section{The quantities \barJk\ and \barJkJl\ for an oblate spheroid}

In order to compare  our results with those of Sorrell (1995a), we
assume that the mantle surface is an oblate spheroid.
If the cosmic ray flux is isotropic in the grain frame,
then the evaporation sites are uniformly distributed
over the grain surface.
The momentum distribution of the evaporating molecules is
determined for thermal evaporation by the principle of
detailed balancing (e.g., Roberge et al.\ 1993, hereafter RDGF93).
Then the required coefficients
are\footnote{In eqs.\ [\ref{2.13}]--[\ref{2.14}], we have neglected terms
of order $\Omega b \over \vvap$, where $\Omega$
is the grain angular velocity and $\vvap$ is
a typical thermal velocity for an evaporating molecule.
Here we have anticipated the result (\S3) that the rotational
energies of the grains are nearly thermal, so that
$\Omega b / \vvap \sim \sqrt{\mvap/M_{\rm d}}$, where
\mvap\ is the mass of an evaporating molecule and $M_{\rm d}$ is
the grain mass.
Note that, although $\barJk=0$ to zeroth order in $\Omega b /\vvap$,
the first-order term we have neglected is always {\it negative}, so
that evaporation from cosmic ray hotspots would
actually spin {\it down}\/ an oblate spheroid.
We will neglect the rotational damping caused by cosmic rays,
which is much smaller than gas damping for reasonable estimates of
the cosmic ray flux.}
\be
\barJk=0, \ \ \ k=x,y,z,
\label{2.13}
\ee
and
\be
\barJkJl= \mvap k\Tvap b^2\epsilon_k\,\delta_{kl} \ \ \ k=x,y,z
\label{2.14}
\ee
(RDGF93, cf.\ eqs.\ [B8] and [B10]), where $b$ is the grain radius
and the components are relative to a Cartesian basis
with \bfzh\ parallel to the grain symmetry axis.
The quantities
\be
\epsilon_k(e) \equiv {4\Gamma_k(e) \over 3 \left[1+(1-e^2)g(e)\right]},
\ \ \ k=x,y,z,
\label{A.16}
\ee
are weak functions of the mantle shape with $1/2 \le \epsilon_k \le 2/3$
and the geometrical factors
\be
\Gamma_z= 
{3 \over 16} \, \left\{\ 
3+4(1-e^2)g(e)-e^{-2}\left[1-(1-e^2)^2g(e)\right]
\right\},
\label{A.17}
\ee
and
\be
\Gamma_x=\Gamma_y = 
{3 \over 32} \, \left\{\ 
7-e^2+(1-e^2)^2g(e)+
(1-2e^2)\left[1+e^{-2}\left[1-(1-e^2)^2g(e)\right]\right]\right\},
\label{A.18}
\ee
where
\be
g(e) \equiv {1 \over 2e} \ln\left({1+e \over 1-e}\right)
\label{A.19}
\ee
are weak functions of $e$ with $3/8 \le \Gamma_k \le 1$.

%
%

%
%
\newpage
\centerline{\bf FIGURE CAPTION}

\noindent
{\bf Fig.~1} ---
The mean number of molecules evaporated by a cosmic ray with energy
loss rate $Q$ equals the number of molecules initially adsorbed in
the effective area $\Aeff(Z,E)$ (see eq.\ [\ref{2.8}]).
Results are shown for the evaporation of \HH\ (solid curves) and
CO (dashed curves).

%
%
\clearpage

\begin{figure}
\begin{picture}(441,216)
\includegraphics{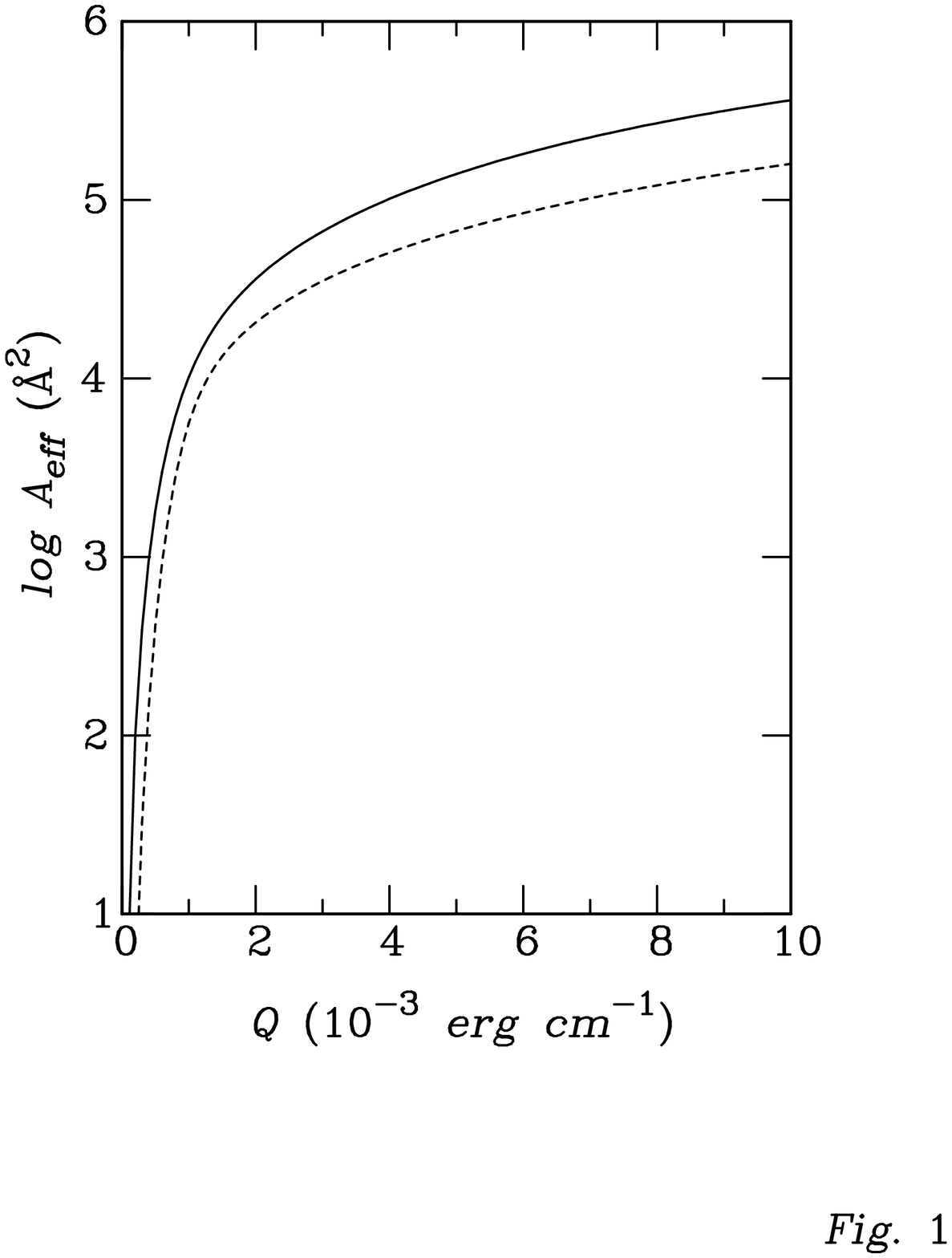}
\end{picture}
\end{figure}


\begin{thebibliography}{}
\bibitem{} Aannestad, P.A., \& Kenyon, S.J. 1979, \apj, 230, 771
\bibitem{} Chrysostomou A., Hough J.H., Whittet, D.C.B., Aitken D.K.,
     Roche P.F., \& Lazarian, A. 1996, \apjl, 465, L61
\bibitem{} Davis, L., \& Greenstein, J.L. 1951, \apj,  114, 206
\bibitem{} DeGraff, T.A., Roberge, W.G., \& Flaherty, J.E. 1997
     (in preparation)
\bibitem{} de Jong, T., \& Kamijo, F. 1973, \aap, 25, 363
\bibitem{} Draine B.T. 1996, in ASP Conf.\ Ser.\ Vol.\ 97, Polarimetry of
     the Interstellar Medium, ed. W.G.\ Roberge \& D.C.B.\ Whittet
     (San Francisco: ASP), 16
\bibitem[]{} Draine, B.T., \& Weingartner J.C. 1996a ApJ, 470, 551.
\bibitem[]{} Draine, B.T., \& Weingartner J.C. 1996b ApJ, submitted
\bibitem{} Jones, R.V., \& Spitzer, L.,Jr 1967, \apj,  147, 943
\bibitem{} Johnson, R.E. 1990, {\it Energetic Charged Particle 
Interactions with Atomspheres and Surfaces}, Springer-Verlag, Berlin, 1990,
p. 112
\bibitem[]{} Lazarian, A. 1995a, MNRAS, 277, 1235
\bibitem{} Lazarian, A.  1995b, \apj, 453, 229
\bibitem{} Lazarian, A., \& Draine B.T. 1996, \apj\ (submitted)
\bibitem{} Lazarian, A. \& Roberge W.G. 1996 \apj\ (submitted)
\bibitem{} Lazarian, A., Goodman, A.A., \& Myers, P.C. 1996, \apj\
     (in preparation)
\bibitem{} L\'{e}ger, A., Jura, M., \& Omont, A. 1985, \aap, 144, 147 (LJO85)
\bibitem{} Purcell, E.M. 1979, \apj, 231, 404 (P79)
\bibitem{} Purcell, E.M. \& Spitzer, L., Jr.\ 1971, \apj,  167, 
\bibitem{} Roberge, W.G., DeGraff, T.A., \& Flaherty, J.E. 1993, \apj, 
     418, 287 (RDGF93)
\bibitem{} Salpeter, E.E., \& Wickramasinghe, N.C. 1969, Nature, 222, 442
\bibitem{} Sandford, S.A., \& Allamandola, L.J. 1993, \apjl, 409, L65
\bibitem{} Simpson, J.A. 1983, \araa, 33, 323
\bibitem{} Sorrell, W.H. 1995a, \mnras, 273, 169
\bibitem{} Sorrell, W.H. 1995b, \mnras, 273, 187
\bibitem{} Watson, W.D., \& Salpeter, E.E. 1972, \apj, 174, 321
\bibitem[]{} Spitzer L.,Jr, \& McGlynn T.A. 1979, {\it ApJ}, {\bf 231}, 417
\bibitem{} Zeller, R.C., \& Pohl, R.O. 1971, \prb, B4, 2029
\end{thebibliography}
\end{document}